\documentclass[fleqn,12pt,twoside]{article}
\usepackage{espcrc1}

\usepackage{graphicx}

\newcommand{\AmS}{{\protect\the\textfont2
  A\kern-.1667em\lower.5ex\hbox{M}\kern-.125emS}}

\title{Deducing the asymptotic normalization constant \\
of the $2^+$ subthreshold state in $^{16}$O from $^{12}$C + $\alpha$ elastic scattering}

\author{J.-M.\ Sparenberg\address{TRIUMF Theory Group, \\
4004 Wesbrook Mall, Vancouver, British Columbia, Canada V6T 2A3}}
      
\begin{document}

\maketitle

\begin{abstract}
R-matrix analyses of the $^{12}$C + $\alpha$ elastic-scattering phase shifts deduced from a 
recent high-precision measurement of the differential cross sections are performed. The $\ell=0$ 
phase shifts constrain the R-matrix radius $a$ around 5.85 fm, while the $\ell=2$ phase shifts lead to a 
strong constrain neither on $a$ nor on the asymptotic normalization constant $C$ of the $2^+$ 
subthreshold state (except for a loose upper limit). This contradicts previous R-matrix analyses of 
the $^{12}$C + $\alpha$ elastic scattering and explains the incompatibility between values of 
$C$ obtained in these analyses.
\end{abstract}

\section{Introduction}

The $^{12}$C $(\alpha, \gamma)$ $^{16}$O capture reaction is very important in nuclear 
astrophysics: combined with the triple-$\alpha$ reaction, it determines the rate of carbon and 
oxygen resulting from the helium-burning phase in red giants. Measuring this reaction rate is a 
difficult task: due to the Coulomb barrier, the cross section is very small ($10^{-17}$ barns at 
$E_\mathrm{c.m.}=300$ keV, the astrophysically-relevant center-of-mass energy). Hence, a theoretical 
extrapolation from measured higher energies is necessary, which is generally performed with the R-matrix formalism.
However, this extrapolation is made very unstable by the presence of two 
bound states in the $^{16}$O spectrum just below the $^{12}$C + $\alpha$ threshold: a $1^-$ 
state strongly influences the dominant E1 capture, while a $2^+$ state strongly affects the 
smaller E2 component. Despite some controversy \cite{gai:98},
the E1 component is generally believed to be 
well constrained by the measurement of the $\beta$-delayed $\alpha$ spectrum of $^{16}$N
\cite{azuma:94}, 
leaving the influence of the $2^+$ subthreshold state as the dominant source of uncertainty on 
the total reaction rate.

In the R-matrix formalism, the key parameter is the bound-state asymptotic normalization 
constant (ANC), which is related to the reduced width. Since this quantity also appears in the 
description of elastic scattering, it has been proposed to constrain it by precise measurement of 
elastic-scattering differential cross sections. An R-matrix analysis of the $2^+$ phase shifts 
deduced from an older experiment \cite{plaga:87}
leads to the value $C=402 \times 10^3$ fm$^{-1/2}$, but no 
error bar is provided \cite{angulo:00}.
A recent high-precision measurement of $N=402$ differential cross sections
at center-of-mass energies between 2 and 6.5 MeV, motivated by the 
astrophysical context described above, has been performed at Notre-Dame \cite{tischhauser:02}.
Though the experimental results are compatible with the older 
experiment, their R-matrix analysis leads to a totally different ANC value: $C=154 \pm 18 
\times 10^3$ fm$^{-1/2}$. This discrepancy suggests a problem in these R-matrix analyses, which 
motivates the present work.

In the next section, I perform new R-matrix analyses of the phase shifts deduced from the Notre-Dame 
experiment \cite{buchmann:03} and solve the above contradiction.
A brief comparison with the potential model is then made in the last section.

\section{R-matrix analyses of the $^{12}$C + $\alpha$ elastic-scattering phase shifts}

In the R-matrix description of elastic scattering, a partial-wave decomposition of the differential 
cross section is performed. The phase shifts of a given partial wave are expressed as the sum of a 
hard-sphere background term with radius $a$ and an R-matrix or "resonant" term which is 
parameterized in terms of energies and reduced widths of the states in the given partial 
wave. These states may be actual resonances at positive energies, which lead to rapid variations in 
the elastic-scattering excitation functions and are experimentally visible, but they may also lie 
outside the measured energy region and only lead to a smooth background influence in the 
measured region. These background terms may correspond either to bound states (negative 
energy), which have a clear physical meaning, or to high-(positive-)energy states, which are 
generally introduced for phenomenological reasons and have a less clear physical meaning.

\begin{figure}[ht]
\centerline{\includegraphics[scale=0.6]{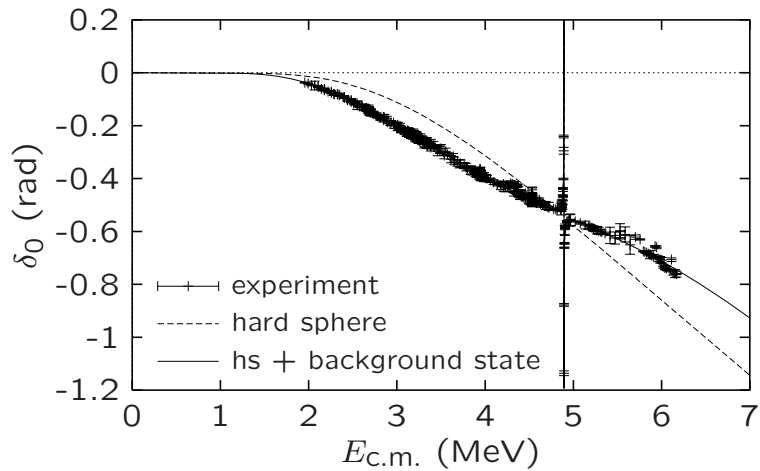}}
\caption{\label{fig:d0hshep}
$\ell=0$ phase shifts as a function of the center-of-mass energy:
experimental points \cite{buchmann:03},
R-matrix fit with a background consisting of (i) a hard-sphere term only (dashed line)
and (ii) a hard-sphere and a high-energy-state terms (full line).}
\end{figure}

\begin{figure}[ht]
\centerline{\includegraphics[scale=0.6]{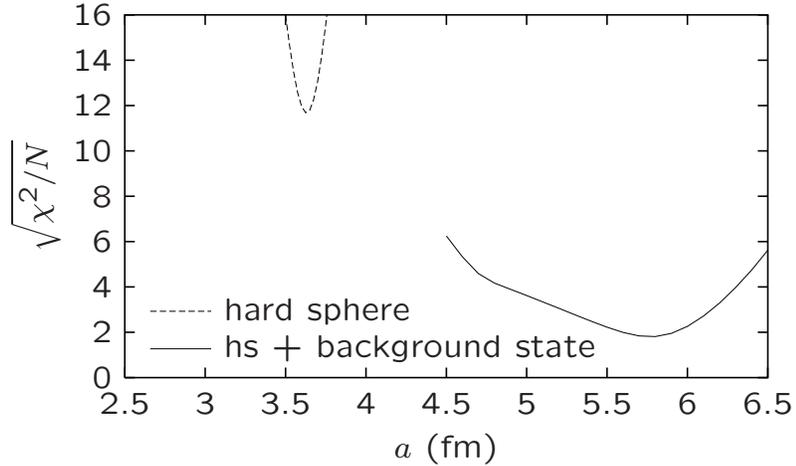}}
\caption{\label{fig:chi0hshep}
$\sqrt{\chi^2/N}$ on the $N=402$ $\ell=0$ phase shifts from Ref.\ \cite{buchmann:03}
for an R-matrix fit with a background consisting of (i) a hard-sphere term only (dashed line)
and (ii) a hard-sphere and a high-energy-state terms (full line),
as a function of the R-matrix radius.}
\end{figure}

Figure \ref{fig:d0hshep} shows a fit of the $\ell=0$ $^{12}$C + $\alpha$ elastic-scattering
phase shifts. 
There is only one narrow resonance at 4.9 MeV, superimposed on a smooth background. Let us 
examine the different contributions to that background by first considering a simple hard-sphere 
term. The dashed curve is the best fit, obtained for $a=3.66$ fm; it corresponds to a deep 
minimum of $\chi^2$ as a function of $a$, as shown on Fig.\ \ref{fig:chi0hshep}, but it is clearly not 
satisfactory. The first way to improve this fit is to consider bound states. I have checked that the 
two $0^+$ bound states of $^{16}$O, when added into the fit, do not improve the situation: 
they are too far away from the measured region. In Ref.\ \cite{tischhauser:02}, the 
$0_2^+$ state is however taken into account into the fit and a very small reduced width (and 
hence a very small ANC) is obtained. I believe this small ANC has no physical meaning: 
it only indicates that the fit is not less good when the bound state is not included. The second 
way to improve the fit is to consider a high-energy phenomenological state. The full curves in 
Figs.\ \ref{fig:d0hshep} and \ref{fig:chi0hshep} correspond to that case.
The fit is now satisfactory; it corresponds to a clear 
minimum of $\chi^2$ for $a=5.85$ fm, which is close to the value of $5.5$ fm obtained in 
Ref.\ \cite{tischhauser:02}.

The present analysis thus confirms that the R-matrix radius can be constrained by elastic 
scattering, as claimed in Ref.\ \cite{tischhauser:02}. The origin of this constrain is mostly the 
$\ell=0$ partial wave: I have checked that the minimum of $\chi^2$ is much less marked for 
other partial waves (see below the case of $\ell=2$). This can be explained by the fact that the 
background phase shifts decrease for increasing angular momentum. The present results also 
show that the value of the radius strongly depends on the presence or absence of a high-energy 
background state. While this radius is physically understandable in the absence of high-energy 
state (3.66 fm is close to the sum of the matter radii of the $^{12}$C and $\alpha$ nuclei), the fit 
is unsatisfactory; on the other hand, with a high-energy state, the fit becomes good but the 
physical meaning of the radius, 5.85 fm, is less clear.

Let us now go to higher angular momenta: for each partial wave, a high-energy state is necessary 
to get a satisfactory fit of the phase shifts, as for $\ell=0$. The background contains thus at least 
two terms. To extract a bound-state ANC from the phase shifts, one has thus to disentangle the 
bound-state background contribution from a total background which consists of three terms (see 
Fig.\ 2 of Ref.\ \cite{angulo:00}). I have shown above that this was not possible for $\ell=0$. 
The same conclusion holds for the $3^-$ bound state, in contrast to what is done in Ref.\ 
\cite{tischhauser:02}: the fit of the phase shifts is not better with the bound state. In both cases, 
this can be understood by the rather low energy of the bound states;
for subthreshold states, one 
could expect the situation to be better. For the $1^-$ state, however, no evidence of better fit is 
obtained in Ref.\ \cite{tischhauser:02}. The general tendency thus seems to be that bound states 
do not make the fit of elastic scattering better. With this in mind, the results of Refs.\ 
\cite{angulo:00} and \cite{tischhauser:02} for $\ell=2$ are rather surprising.

\begin{figure}[ht]
\centerline{\includegraphics[scale=0.6]{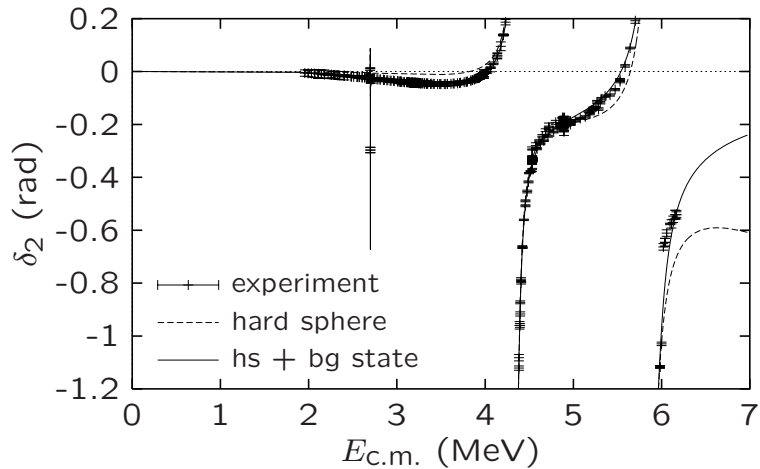}}
\caption{\label{fig:d2hshep}
Same as Fig.\ \ref{fig:d0hshep} but for $\ell=2$.}
\end{figure}

\begin{figure}[ht]
\centerline{\includegraphics[scale=0.9]{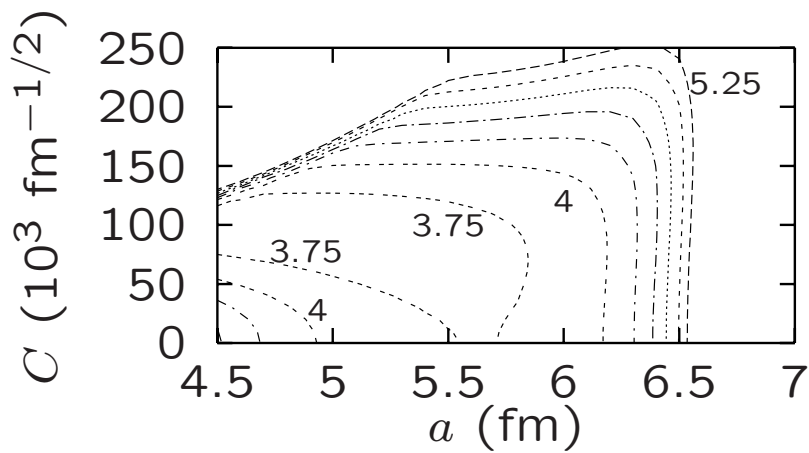}}
\caption{\label{fig:chi2hsbshep}
$\sqrt{\chi^2/N}$ on the $N=402$ $\ell=2$ phase shifts from Ref.\ \cite{buchmann:03}
for an R-matrix fit with a background consisting a hard-sphere,
a high-energy-state and a bound-state terms,
as a function of the R-matrix radius $a$ and the bound-state ANC $C$.
The best fits correspond to $\sqrt{\chi^2/N}\approx 3.7$.}
\end{figure}

I have tried to reproduce these results but did not find any relevant $\chi^2$ minimum 
which could lead to a constrain on the ANC.
The situation for $\ell=2$ is actually very similar to that for $\ell=0$, as shown by Fig.\
\ref{fig:d2hshep}:
a good fit requires a high-energy background state but does not require a bound-state term.
The situation is best illustrated on Fig.\ \ref{fig:chi2hsbshep}, where the $\chi^2$ on the phase shifts
is plotted as a function of $a$ and $C$.
There is a valley of satisfactory values for these parameters,
which ranges from $C=0$ (no bound state in the fit) to at least $C=200 \times 10^3$ fm$^{-1/2}$.
These fits are satisfactory in the sense that they give a good graphical fit of the phase shifts 
(see Fig.\ \ref{fig:d2hshep})
but they are not totally satisfactory from the point of view of $\chi^2$ theory,
which provides me from rigorously defining the acceptable values of $C$.
This is probably due to an underestimation of systematic errors on the phase shifts of Ref.\ \cite{buchmann:03}.
Figure \ref{fig:chi2hsbshep} however clearly shows that the R-matrix fit only provides an upper limit
on the ANC, strongly depending on $a$.
The conclusion is thus the same for all partial waves:
an R-matrix analysis of a single partial wave is not able to constrain a bound-state ANC,
even when the bound state is close to threshold.
This shows that the disagreeing values of $C$ obtained in Refs.\ \cite{angulo:00} and \cite{tischhauser:02} are unreliable.
I think that the $\chi^2$ minimum obtained in Ref.\ \cite{angulo:00} is due to too restrictive hypotheses on the R-matrix radius and the high-energy state, and that the error bar on $C$ obtained in Ref.\ \cite{tischhauser:02} neglects systematic errors on the data and is therefore much too small.

\section{Comparison with the potential model}

The above result is not very surprising from the point of view of the potential model:
it is well known that, for a given partial wave,
different potentials may have bound states at the same energy but different ANC,
for the same scattering phase shifts at all energies.
Such {\em phase-equivalent} potentials are called Bargmann potentials
\cite{bargmann:49a,bargmann:49b}
and are well known in the context of the inverse problem \cite{chadan:77}.
Such potentials have been constructed for $^{12}$C + $\alpha$ in Ref.\ \cite{sparenberg:04}.
In the same reference, it is shown that the ANC may however be extracted from the elastic-scattering
phase shifts, in the framework of the potential model,
provided several partial waves are considered simultaneously.
For $^{12}$C + $\alpha$, this can be achieved by fitting the properties of the $0_1^+$
rotational band, to which the $2^+$ subthreshold state belongs.
This leads to the value $C=144.5 \pm 8.5 \times 10^3$ fm$^{-1/2}$.


\end{document}